\documentclass[preprint]{revtex4}

\usepackage{graphicx}

\begin{document}

\title{The Power of Axisymmetric Pulsar}

\author{Andrei Gruzinov}
 
\affiliation{Center for Cosmology and Particle Physics, Department of Physics, New York University, NY 10003}

\date{July 13, 2004}

\begin{abstract}

Stationary force-free magnetosphere of an axisymmetric pulsar is shown to have a separatrix inclination angle of 77.3$^\circ$. The electromagnetic field has an $R^{-1/2}$ singularity inside the separatrix near the light cylinder. A numerical simulation of the magnetosphere which crudely reproduces these properties is presented. The numerical results are used to estimate the power of an axisymmetric pulsar: $L=(1\pm 0.1)\mu^2\Omega^4/c^3$. A need for a better numerical simulation is pointed out.

\end{abstract}

\maketitle

\section{Introduction}

A neutron star with magnetic dipole $\mu$, rotating around its magnetic axis with frequency $\Omega$ loses energy at a about the magneto-dipole rate $L \sim \mu ^2\Omega^4/c^3$. This is because the star creates free charges that form a magnetosphere with non-zero Poynting flux along open field lines -- the well-known prediction of Goldreich and Julian \cite{gold}. 

The shape of the force-free axisymmetric pulsar magnetosphere was first calculated by Contopoulos, Kazanas, and Fendt \cite{cont}. This important paper demonstrated that a stationary solution does exist, and the power of the pulsar must be close to the magneto-dipole value. 

However, in \S2 by solving the stationary force-free equations in the vicinity of the critical circle (the intersection of the light cylinder and the equatorial plane), we show that (i) the separatrix inclination angle is equal to 77.3$^\circ$, (ii) the electromagnetic field has an $R^{-1/2}$ singularity near the critical circle inside the separatrix. Neither of these properties are seen in the numerical results of \cite{cont}. 

We therefore repeated the simulation of \cite{cont} and found the following (\S3). At numerical resolution similar to that of \cite{cont}, our code reproduces most of their results. But at higher resolution, the separatrix steepens and fattens, and the singularity of the electromagnetic field inside the separatrix starts to develop. The numerical simulation crudely reproduces the predicted properties of the magnetosphere. We used our numerical results to estimate the power of an axisymmetric pulsar: $L=(1\pm 0.1)\mu^2\Omega^4/c^3$.

\begin{figure}[b]
  \begin{center}
    \includegraphics[angle=0, width=.4\textwidth]{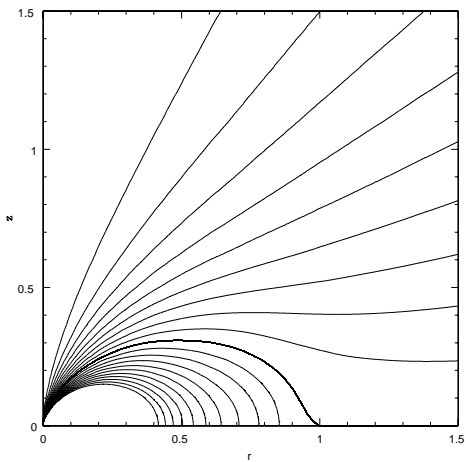}
    \caption{Stationary force-free axisymmetric pulsar magnetosphere. The thick line shows the separatrix $\psi _0=1.27$. Thin lines correspond to $\psi$ intervals of $0.1\psi _0$.}
  \end{center}
\end{figure}

\subsection{The pulsar magnetosphere equation}
In Appendix A we describe force-free electrodynamics -- a remarkable version of plasma physics without any plasma properties appearing explicitly. There we also discuss the applicability of the force-free approximation. To finish the introduction we must give a brief derivation of the pulsar magnetosphere equation \cite{scharl}.

It is assumed that electromagnetic forces are much stronger than inertia, thus ${\bf j}\times {\bf B}+c\rho {\bf E}=0$, or 
\begin{equation}\label{fund}
\nabla \times {\bf B}\times {\bf B}+\nabla ^2\phi \nabla \phi =0
\end{equation}
Here $E$, $B$ are electric and magnetic fields, $\rho$, $j$ are charge and current density, $\phi$ is the electrostatic potential; the fields are stationary. For axisymmetric fields, we represent the magnetic field by the toroidal component of the vector potential $\psi /r$ and by the quantity $A\equiv 2I/c$, where $I$ is the poloidal current: 
\begin{equation}\label{cyl}
{\bf B}=\left( -{\psi _z\over r}, ~{A\over r}, ~{\psi _r\over r}\right),
\end{equation}
in cylindrical coordinates $r,z$; subscripts mean partial derivatives.  Using (\ref{cyl}) in (\ref{fund}), one gets the pulsar magnetosphere equation \cite{scharl}:
\begin{equation}\label{basic}
(1-r^2)(\psi _{rr}+{1\over r}\psi _r+\psi _{zz})-{2\over r}\psi _r+F(\psi )=0.
\end{equation}
here $F(\psi )\equiv A(\psi ){dA(\psi )\over d\psi }$, $A(\psi )$ is an arbitrary function, $\phi =\psi$ follows from the boundary condition on the surface of the star, and we use dimensionless units 
\begin{equation}
c=\Omega =\mu =1.
\end{equation}
The basic equation (\ref{basic}) must be solved with the small distance boundary condition 
\begin{equation}
\psi \rightarrow {r^2\over (r^2+z^2)^{3/2}}, ~~~~~~~r,z\rightarrow 0,
\end{equation}
which corresponds to the dipole field. 

Michel \cite{michel} has found an exact solution of the magnetosphere equation for a magnetic monopole rather than a dipole star, proving that solutions of (\ref{basic}) which are  smooth across the light cylinder ($r=1$) do exist in some cases.

\section{Near the singular circle}

The basic equation (\ref{basic}) can be solved in the vicinity of the singular circle, $|z|\ll 1$ and $|r-1|\ll 1$. In this region, (\ref{basic}) can be approximated as 
\begin{equation}\label{basic1}
x(\psi _{xx}+\psi _{zz})+\psi _x={1\over 2}F,
\end{equation}
where $x\equiv r-1$. 

We assume (to be confirmed by numerical simulations of \S3) that there is a nonzero return current flowing along the separatrix (Fig.1 ). Let $\psi _0$ be the value of the potential on the separatrix, and $A_0\equiv A(\psi_0 -0)$. Then the return current is equal to $A_0/2$. From (\ref{basic1}), we get the jump condition across the separatrix  
\begin{equation}
(\nabla \psi)^2|_{\psi =\psi _0+0}-(\nabla \psi)^2|_{\psi =\psi _0-0}=-{1\over 2x}A_0^2.
\end{equation}
In the closed line region, for $\psi >\psi _0$, we must therefore have $\psi \propto (-x)^{1/2}$. Thus electric and magnetic fields diverge as inverse square root in the vicinity of the singular circle. This is an admissible singularity, since the total energy of the fields remains finite. 

We can now find the leading order solution in the closed line region. We set 
\begin{equation}
\psi = \psi _0+R^{1/2}f(\theta),
\end{equation}
where $x\equiv R\sin \theta$ and $z\equiv R\cos \theta$. Then (\ref{basic1}) gives inside the separatrix
\begin{equation}
{1\over \sin \theta}{d\over d\theta}\left(\sin \theta {df\over d\theta}\right)+{3\over 4}f=0.
\end{equation}
Solving this ordinary equation numerically for $df/d\theta =0$ at $\theta =-\pi /2$, we find $f(\theta =-0.222)=0$. Thus the inclination angle of the separatrix is 77.3$^\circ$.

Knowing the separatrix inclination angle, we can solve (\ref{basic1}) in the open line region too. We assume that in the open line region $\psi -\psi _0=R^{\alpha}f(\theta )$, and correspondingly $F \propto (\psi _0-\psi)^{1-{1\over \alpha }}$. Then equation (\ref{basic1}) reads
\begin{equation}
{1\over \sin \theta}{d\over d\theta}\left(\sin \theta {df\over d\theta}\right)+\alpha (\alpha +1)f={C\over \sin \theta}f^{1-{1\over \alpha }}.
\end{equation}
Here the two free parameters $\alpha$ and $C$ should be adjusted so as to have (i) $f(-0.222)=0$, (ii) $f(\pi /2) =0$, (iii) no singularity at $\theta =0$. Numerical solution gives
\begin{equation}\label{outer}
\psi _0 -\psi  \propto R^{2.4}, ~~~~~~~~~~~~~F(\psi ) \propto -(\psi _0-\psi)^{0.58}.
\end{equation}

All these properties are roughly reproduced by the numerical simulation presented in \S3.

\begin{figure}[b]
  \begin{center}
    \includegraphics[angle=0, width=.4\textwidth]{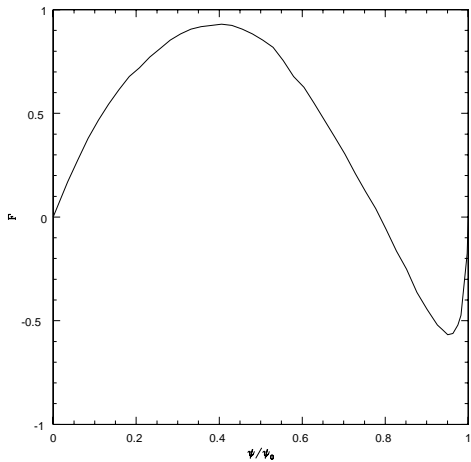}
    \caption{The function $F$ which makes the solution of the magnetosphere equation (\ref{basic}) smooth across the light cylinder.}
  \end{center}
\end{figure}

\section{The axisymmetric pulsar magnetosphere}
Numerical solution of the axisymmetric pulsar equation (\ref{basic}) can be obtained in the following way \cite{cont,ogura}. One takes an arbitrary $F(\psi )$ and solves (\ref{basic}) in the inner ($r<1$) and in the outer ($r>1$) regions using appropriate boundary conditions (the boundary condition at the light cylinder being $\psi _r=F(\psi )/2$). This, of course, does not give the true solution, because one gets $\psi (1-0,z)\neq \psi(1+0,z)$. One then makes a number of adjustments of $F(\psi )$, aimed at reducing the jump $\psi (1-0,z)- \psi(1+0,z)$ for all $z$. An important finding of \cite{cont} is that this procedure actually gives an everywhere smooth solution. 

In our simulation, we followed the same method. We were adjusting $F(\psi )$ until an acceptable solution was obtained. A solution was called acceptable if the following integral $\int _0 ^{\infty} dz (\psi (1-0,z)- \psi(1+0,z))^2/(1+z^2)$ was reduced to less then $10^{-7}$ (starting from $\approx 1$ at $F\equiv 0$). Equation (\ref{basic}) was solved by a simple relaxation method. Simultaneously with the relaxation, the adjustment of $F$ was carried out, in a way similar to that of \cite{cont}. After an acceptable solution was obtained, the adjustment of $F$ was stopped, while the relaxation was carried out for a sufficient number of steps to ensure that our $\psi(r,z)$ does solve (\ref{basic}) for the obtained $F(\psi)$.

The full function $F_{\rm full} $ which must be used in (\ref{basic}) consists of the regular piece $F$ (the one shown in Fig.2 ), and the delta function piece $F_{\delta}(\psi )=-(\int F)\delta (\psi -\psi _0)$. In the numerical simulation the delta-function piece was smoothed over the $\psi$ interval $(\psi_0, \psi_0(1+d))$. The simulation was carried out for different values of $d$. The figures show the case $d=0.03$. The table lists some of the simulations that were carried out. The independence of numerical results on the radius of the star, for small radii of the star, and on the outer boundaries location, for distant boundaries,  was checked.

\begin{table}
\caption{Simulation Results}
\begin{ruledtabular}
\begin{tabular}{lllll}
Resolution & $\delta$-function width $d$ & Separatrix value $\psi_0$ & Separatrix inclination, $^\circ$ & Power \\

$200\times 100$  & 0.02  & 1.27 & 63 &  1.01\\
$200\times 100$  & 0.03  & 1.27 & 60 &  1.02\\
$200\times 100$  & 0.04  & 1.30 & 56 &  1.06\\
$200\times 100$  & 0.06  & 1.32 & 50 &  1.10\\
$200\times 100$  & 0.08  & 1.35 & 48 &  1.15\\
$200\times 100$  & 0.1  & 1.38 & 48 &  1.19\\
$100\times 50$  & 0.1  & 1.40 & 42 &  1.27\\
$100\times 50$  & 0.06  & 1.34 & 48 &  1.17\\

\end{tabular}
\end{ruledtabular}
\end{table}

The power of the pulsar, which is proportional to the spin-down rate, is obtained by integrating the Poynting flux over an arbitrary sphere. One gets the dimensionless power
\begin{equation}
L=\int _0^{\psi _0}d\psi A(\psi ),
\end{equation}
where $A(\psi)=\left( 2\int _0^{\psi }d\psi 'F(\psi ')\right)^{1/2}$ is obtained from the regular part of $F$. From the table, we estimate the power of an axisymmetric pulsar $L=(1\pm 0.1)\mu^2\Omega^4/c^3$.

Our numerical solution reproduces the features predicted in \S2 in the following sense: (i) the inclination angle increases with decreasing $d$ and the extrapolated value of inclination is about 70$^\circ$ (ii) the maximum of the magnetic field in the inner part of the separatrix becomes more pronounced with decreasing $d$, (iii) the function $F$ demonstrates a singularity similar to (\ref{outer}).

However, we have not accurately reproduced either of the 4 numbers given \S2. A really good numerical solution should show: (i) the 77$^\circ$ separatrix inclination, (ii) the $\psi -\psi _0\propto R^{1/2}$ singularity in the inner region, (iii) the $F(\psi ) \propto -(\psi _0-\psi)^{0.58}$ singularity, (iv) the $\psi _0 -\psi  \propto R^{2.4}$ singularity in the outer region. Until such a solution is obtained, one cannot be really sure that a stationary force-free pulsar magnetosphere exists, and one cannot be really sure that the power of an axisymmetric pulsar is $L=(1\pm 0.1)\mu^2\Omega^4/c^3$.

\begin{acknowledgments}
This work was supported by the David and Lucile Packard Foundation.
\end{acknowledgments}

\begin{appendix}

\section{Force-free electrodynamics (FFE) }

(This Appendix contains a large excerpt from astro-ph/9902288)
FFE is applicable if electromagnetic fields are strong enough to produce pairs and baryon contamination is prevented by strong gravitational fields \cite{blan}. Pulsars, Kerr black holes in external magnetic fields, relativistic accretion disks, and gamma-ray bursts are the astrophysical objects whose luminosity might come originally in a pure electromagnetic form describable by FFE. 

FFE is classical electrodynamics supplemented by the force-free condition:
\begin{equation}
\partial _t{\bf B}=-\nabla \times {\bf E},
\end{equation}
\begin{equation}
\partial _t{\bf E}=\nabla \times {\bf B}-{\bf j},
\end{equation}
\begin{equation}
\rho {\bf E}+{\bf j}\times {\bf B}=0.
\end{equation}
$\nabla \cdot {\bf B}=0$ is the initial condition. The speed of light is $c=1$; $\rho =\nabla \cdot {\bf E}$ and ${\bf j}$ are the charge and current densities multiplied by $4\pi$. The electric field is everywhere perpendicular to the magnetic field, ${\bf E}\cdot {\bf B}=0$. The electric field component parallel to the magnetic field should vanish because charges are freely available in FFE. It is also assumed that the electric field is everywhere weaker than the magnetic field, $E^2<B^2$. Then equation (A3) means that it is always possible to find a local reference frame where the field is a pure magnetic field, and the current is flowing along this field. FFE is Lorentz invariant. 

Equation (A3) can be written in the form of the Ohm's law. The current perpendicular to the local magnetic field can be calculated from equation (A3). The parallel current is determined from the condition that electric and magnetic fields remain perpendicular during the evolution described by the Maxwell equations (A1), (A2).  We thus obtain the following non-linear Ohm's law
\begin{equation}
{\bf j}={({\bf B}\cdot \nabla \times {\bf B}-{\bf E}\cdot \nabla \times {\bf E}){\bf B}+(\nabla \cdot {\bf E}){\bf E}\times {\bf B} \over B^2}.
\end{equation}
Equations (A1), (A2), (A4) form an evolutionary system (initial condition ${\bf E}\cdot {\bf B}=0$ is assumed). It therefore makes sense to study stability of equilibrium electromagnetic fields in FFE. One can also study linear waves and their nonlinear interactions in the framework of FFE \cite{blae}.

One can introduce a formulation of FFE similar to magnetohydrodynamics (MHD); then we can use the familiar techniques of MHD to test stability of magnetic configurations. To this end, define a field ${\bf v}={\bf E}\times {\bf B}/B^2$, which is similar to velocity in MHD. Then ${\bf E}=-{\bf v}\times {\bf B}$ and equation (A1) becomes the ``frozen-in'' law
\begin{equation}
\partial _t{\bf B}=\nabla \times ({\bf v}\times {\bf B}).
\end{equation}
From ${\bf v}={\bf E}\times {\bf B}/B^2$, and from equations (A1)-(A3), one obtains the momentum equation
\begin{equation}
\partial _t(B^2{\bf v})=\nabla \times {\bf B}\times {\bf B}+\nabla \times {\bf E}\times {\bf E}+(\nabla \cdot {\bf E}){\bf E},
\end{equation}
where ${\bf E}=-{\bf v}\times {\bf B}$. Equations (A5), (A6) are the usual MHD equations except that the density is equal $B^2$ and there are order $v^2$ corrections in the momentum equation.

We must mention that the applicability of FFE to pulsars can be questioned \cite{mic}. If charges are not freely available, some regions of the pulsar magnetosphere might exist that should be described by a vacuum rather than force-free electrodynamics. While we cannot offer a real description of the creation of the space charge, a simple energy estimate shows that the system might find a way to put the entire magnetosphere into the force-free regime. Indeed, the pulsar luminosity is $L \sim B^2R^6\Omega ^4/c^3$, where $B$ is the magnetic field, $R$ is the radius. The number density of charged particles is $n \sim  \Omega B/(ce)$. The associated energy density is $nmc^2$, and the associated power in particles is $L_p\sim nmc^3R^2$. The ratio $L_p/L\sim mc^5/(e\Omega ^3BR^4)$ is a very small number everywhere in the magnetosphere. Thus, energy-wise, charges are indeed freely available. With only a tiny fraction of the pulsar luminosity channeled into the charge production, the star will be able to put the entire magnetosphere into the force-free state. The real mechanisms for populating the magnetosphere are of course of great importance, but these probably include complex interactions of the pulsar radiation, high-energy electrons and positrons, the surface of the neutron star, the large-scale and turbulent electromagnetic fields --- should be difficult to decipher. But FFE might well turn out to be a good approximation to reality. 

\end{appendix}

\end{document}